# Explicit TE/TM scheme for particle beam simulations

Martin Dohlus and Igor Zagorodnov

*Deutsches Elektronen-Synchrotron,, Notkestrasse 85, 22603 Hamburg, Germany*

**Abstract**

In this paper we propose an *explicit* two-level conservative scheme based on a TE/TM like splitting of the field components in time. Its dispersion properties are adjusted to accelerator problems. It is simpler and faster than the implicit version [1]. It does not have dispersion in the longitudinal direction and the dispersion properties in the transversal plane are improved. The explicit character of the new scheme allows a uniformly stable conformal method without iterations and the scheme can be parallelized easily. It assures energy and charge conservation. A version of this explicit scheme for rotationally symmetric structures is free from the progressive time step reducing for higher order azimuthal modes as it takes place for Yee's explicit method used in the most popular electrodynamics codes.

## 1. Introduction

In particle accelerators based on radio frequency or laser technology a preferred direction - direction of motion is well defined. The electromagnetic field changes very fast in this direction and it is extremely important to eliminate the numerical dispersion error in the direction of motion for *all* frequencies. If the numerical dispersion is suppressed then a quite coarse mesh and moderate computational resources can be used to reach accurate results. It was shown, for example, in wakefield calculations by A. Novokhatski [2] and in laser-plasma interaction simulations by A. Pukhov [3].

The simplest solution is to use the conventional Yee's FDTD scheme with the direction of motion aligned along a grid diagonal. This approach was considered, for example, in [4]. To send the bunch or laser pulse along the grid diagonal is quite inconvenient for numerical realization and increases geometrical errors in approximation of cylindrical accelerator elements parallel to the direction of the motion. Another drawback of this approach is the requirement of an equidistant mesh with the same steps in all directions.

An alternative solution is to develop a scheme without dispersion along an axis. Such methods are described in [2,3] for two dimensional and in [1, 5-7] for three dimensional problems. However, all these approaches lose in simplicity, efficiency and memory demands compared with Yee's scheme [8].

In this paper we present a scheme which competes with Yee's algorithm. The scheme does not have dispersion in the axis direction. It is based on a TE/TM ("transversal electric - transversal magnetic") like splitting of the field components in time. It is simpler and faster than the implicit version, introduced earlier in [1]. The numerical effort is scaled as 5/3 compared to Yee's algorithm for the same time step. But the explicit scheme allows a larger time step than the Yee's algorithm. With such choice the explicit TE/TM scheme requires only ~18% more computational time. The memory demands are the same. The explicit character of the new scheme allows for the uniformly stable conformal method [9] to reach the second order convergence and the scheme can be parallelized easily.

A version of this explicit scheme for rotationally symmetric structures is free from the progressive time step reducing for higher order azimuthal modes as it takes place for conventional Yee's Finite Difference Time Domain (FDTD) method used in the most popular accelerator codes [10].

## 2. Space discretization and matrix notation

In the following we consider a Causchy problem for the Maxwell's equations

$$\nabla \times \vec{H} = \frac{\partial}{\partial t}\vec{D} + \vec{j}, \qquad \nabla \times \vec{E} = -\frac{\partial}{\partial t}\vec{B}, \qquad (1)$$

$$\nabla \cdot \vec{D} = \rho, \qquad \nabla \cdot \vec{B} = 0$$

$$\vec{H} = \mu^{-1}\vec{B}, \qquad \vec{D} = \varepsilon\vec{E}, \qquad x \in \Omega.$$

$$\vec{E}(t=0) = \vec{E}_0, \qquad \vec{H}(t=0) = \vec{H}_0, \qquad x \in \bar{\Omega},$$

$$\vec{n} \times \vec{E} = 0, \qquad x \in \partial\Omega,$$

where $\vec{E}_0, \vec{H}_0$ is an initial electromagnetic field in the domain $\bar{\Omega} = \Omega \cup \partial\Omega$ with boundary $\partial\Omega$.

It follows from (1) that the continuity equation holds

$$\frac{\partial}{\partial t}\rho + \nabla \cdot \vec{j} = 0 \qquad (2)$$

and the energy law is fulfilled

$$\frac{\partial}{\partial t}w = \nabla \cdot \left(\vec{E} \times \vec{H}\right) - \vec{j} \cdot \vec{E}, \quad w = 0.5\left(\vec{E} \cdot \vec{D} + \vec{B} \cdot \vec{H}\right). \tag{3}$$

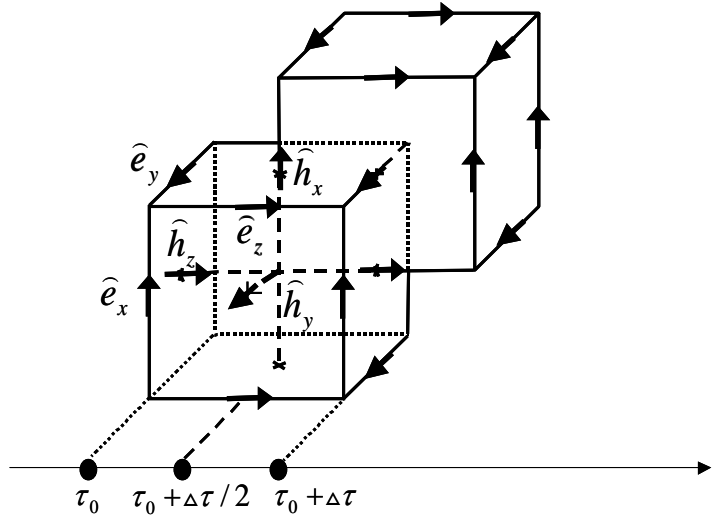

Fig. 1. Positions of a relativistic charged particle in the FIT grid in different moments of time. The scaled time step is chosen equal to the longitudinal mesh step.

Following the matrix notation of the finite integration technique (FIT) [11], the Cauchy problem can be approximated by the *time-continuous* matrix equations on a grid doublet $(G, \tilde{G})$

$$\mathbf{C}\widehat{\mathbf{e}} = -\frac{d}{dt}\widehat{\widehat{\mathbf{b}}}, \quad \mathbf{C}^T\widehat{\mathbf{h}} = \frac{d}{dt}\widehat{\widehat{\mathbf{d}}} + \widehat{\widehat{\mathbf{j}}}, \tag{4}$$

$$\mathbf{S}\widehat{\widehat{\mathbf{b}}} = \mathbf{0}, \quad \tilde{\mathbf{S}}\widehat{\widehat{\mathbf{d}}} = \mathbf{q},$$

completed by the discrete form of the material relations (constitutive equations)

$$\widehat{\mathbf{e}} = \mathbf{M}_{\varepsilon^{-1}}\widehat{\widehat{\mathbf{d}}}, \quad \widehat{\mathbf{h}} = \mathbf{M}_{\mu^{-1}}\widehat{\widehat{\mathbf{b}}},$$

with the discrete inverse permittivity matrix $\mathbf{M}_{\varepsilon^{-1}}$ and the inverse permeability matrix $\mathbf{M}_{\mu^{-1}}$. In the following the material matrices are assumed to be real and symmetric.

On Cartesian $\{x, y, z\}$-coordinate grids (like the Cartesian grid shown in Fig. 1) with an appropriate indexing scheme the curl matrix has a 3x3 block structure [11]:

$$\mathbf{C} = \begin{pmatrix} \mathbf{0} & -\mathbf{P}_z & \mathbf{P}_y \\ \mathbf{P}_z & \mathbf{0} & -\mathbf{P}_x \\ -\mathbf{P}_y & \mathbf{P}_x & \mathbf{0} \end{pmatrix}, \mathbf{S} = \begin{pmatrix} \mathbf{P}_x & \mathbf{P}_y & \mathbf{P}_z \end{pmatrix}, \tilde{\mathbf{S}} = \begin{pmatrix} -\mathbf{P}_x^T & -\mathbf{P}_y^T & -\mathbf{P}_z^T \end{pmatrix}.$$

With changing of variables $\mathbf{e} = \mathbf{M}_{\varepsilon^{-1}}^{-1/2}\widehat{\mathbf{e}}$, $\mathbf{h} = \mathbf{M}_{\mu^{-1}}^{-1/2}\widehat{\mathbf{h}}$, $\mathbf{j} = c^{-1}\mathbf{M}_{\varepsilon^{-1}}^{1/2}\widehat{\widehat{\mathbf{j}}}$, $\tau = ct$, system (4) reduces to the skew-symmetric one

$$\frac{d}{d\tau}\mathbf{e} = \mathbf{C}_0^T\mathbf{h} + \mathbf{j}, \quad \frac{d}{d\tau}\mathbf{h} = -\mathbf{C}_0\mathbf{e}, \quad \mathbf{S}_h\mathbf{h} = \mathbf{0}, \quad \mathbf{S}_e\mathbf{e} = \mathbf{q}. \tag{5}$$

with new "curl" and "div" operators

$$\mathbf{C}_0 = c^{-1}\mathbf{M}_{\mu^{-1}}^{1/2}\mathbf{CM}_{\varepsilon^{-1}}^{1/2} = \begin{pmatrix} 0 & -\mathbf{P}_z^0 & \mathbf{P}_y^0 \\ \mathbf{P}_z^1 & 0 & -\mathbf{P}_x^0 \\ -\mathbf{P}_y^1 & \mathbf{P}_x^1 & 0 \end{pmatrix}.$$

$$\mathbf{S}_e = \begin{pmatrix} -\mathbf{P}_x^T\mathbf{M}_{\varepsilon_x^{-1}}^{-1/2} & -\mathbf{P}_y^T\mathbf{M}_{\varepsilon_y^{-1}}^{-1/2} & -\mathbf{P}_z^T\mathbf{M}_{\varepsilon_z^{-1}}^{-1/2} \end{pmatrix} = \begin{pmatrix} \mathbf{P}_x^e & \mathbf{P}_y^e & \mathbf{P}_z^e \end{pmatrix},$$

$$\mathbf{S}_h = \begin{pmatrix} \mathbf{P}_x\mathbf{M}_{\mu_x^{-1}}^{-1/2} & \mathbf{P}_y\mathbf{M}_{\mu_y^{-1}}^{-1/2} & \mathbf{P}_z\mathbf{M}_{\mu_z^{-1}}^{-1/2} \end{pmatrix} = \begin{pmatrix} \mathbf{P}_x^h & \mathbf{P}_y^h & \mathbf{P}_z^h \end{pmatrix}.$$

System (5) is a *time-continuous* and *space-discrete* approximation of problem (1). As easy to see [11] semi-discrete analogues of conservation laws (2) and (3) hold

$$\frac{d}{d\tau}\mathbf{q} + \mathbf{S}_e\mathbf{j} = 0, \qquad \frac{d}{d\tau}\mathbf{S}_h\mathbf{h} = 0, \qquad (6)$$

$$\frac{d}{d\tau}W = -\langle \mathbf{e}, \mathbf{j} \rangle, \qquad W = 0.5\left[\mathbf{e}^T\mathbf{e} + \mathbf{h}^T\mathbf{h}\right].$$

The next step is a discretization of the system in time. The field components can be split in time and the "leap-frog" scheme can be applied. With "electric/magnetic" splitting a well known Yee's scheme [8] will be obtained. In the following we consider alternative TE/TM schemes.

## 3. Dispersion of Yee's scheme

Suggested by K. Yee [8], the E/M ("electric - magnetic") splitting of the field components yields the *explicit* FDTD scheme (E/M scheme)

$$\mathbf{e}^{n+0.5} = \mathbf{e}^{n-0.5} + \Delta\tau\mathbf{C}_0^*\mathbf{h}^n - \Delta\tau\mathbf{j}^n, \qquad (7)$$

$$\mathbf{h}^{n+1} = \mathbf{h}^n - \Delta\tau\mathbf{C}_0\mathbf{e}^{n+0.5},$$

where $\Delta\tau$ is the time step and the update of the electric components is shifted by $0.5\Delta\tau$ relative to the update of the magnetic components.

On equidistant mesh the scheme has second order local approximation error in homogeneous regions, $O(\|\Delta\vec{r}\|^2 + \Delta\tau^2)$, $\Delta\vec{r} = (\Delta x, \Delta y, \Delta z)^T$. The scheme is stable in vacuum if the next relation [12]

$$\Delta\tau \leq \left(\Delta x^{-2} + \Delta y^{-2} + \Delta z^{-2}\right)^{-0.5}. \qquad (8)$$

on time step holds.

However, the phase velocity of discrete wave modes can differ from the light velocity by an amount varying with the wavelength, direction of propagation in the grid and grid discretization.

The dispersion relation of this scheme in free space has the form [12]

$$\frac{\sin^2\Omega}{\Delta\tau^2} = \frac{\sin^2 K_z}{\Delta z^2} + \frac{\sin^2 K_x}{\Delta x^2} + \frac{\sin^2 K_y}{\Delta y^2}, \qquad (9)$$

where $\Omega = 0.5\omega\Delta\tau/c$, $K_x = 0.5k_x\Delta x$, $K_y = 0.5k_y\Delta y$, $K_z = 0.5k_z\Delta z$. With an equidistant mesh, $\Delta x = \Delta y = \Delta z$, a homogenous material and the time step equal to the right-hand side of inequality (8), the scheme has zero dispersion along the grid diagonals. Hence, the zero dispersion in a desired direction can be achieved by a rotation of the mesh. However, this approach awakes limitations on discretization: the only reasonable choice in this case is to take equal mesh steps in the all three directions. The next difficulty arises with the attempt to use a conformal method [9].

Let us consider a calculation of electromagnetic fields excited by a Gaussian bunch of RMS width $\sigma$ which moves in the z-direction through a structure of length $L$. A self-field of the relativistic bunch has only transverse field components [13] and it is like a plane wave. For the plane wave in z-direction dispersion relation (9) simplifies to

$$\frac{\sin(0.5k\Delta z)}{0.5\Delta z} = \frac{\sin(0.5\Delta\tau\omega/c)}{0.5\Delta\tau}. \quad (10)$$

The numerical wave number $k$ differs from the analytical one by some value $\delta k$ and the Taylor expansion of the last equation up to the first order in $\delta k$ reads

$$\delta k \approx \frac{1}{3!}\left(\frac{\omega}{c}\right)^3\left(\left(\frac{\Delta z}{2}\right)^2 - \left(\frac{\Delta\tau}{2}\right)^2\right). \quad (11)$$

The dispersion error could disappear only when the time step is equal to the mesh step: $\Delta\tau = \Delta z$. But it contradicts stability condition (8). Hence, for any time step the Yee's scheme has a dispersion error in the z-direction of the order

$$\delta k \approx O\left(\left(\frac{\omega}{c}\right)^3 \Delta z^2\right).$$

The Gaussian bunch contains high frequencies up to the frequency $\omega \sim c/\sigma$. Hence, for the structure of length $L$ the phase error will be of the order

$$L \cdot \delta k \approx O\left(L\sigma^{-3}\Delta z^2\right). \quad (12)$$

The last equation means that Yee's scheme [8] demands a very fine mesh for short bunches and long structures with the mesh step

$$\Delta z \ll \sigma^{3/2} L^{-0.5}. \quad (13)$$

In the next section we introduce a scheme without dispersion error in the z-direction. For this scheme the mesh step is independent from the structure length $L$ and is related only to the first power of the bunch length

$$\Delta z \ll \sigma. \quad (14)$$

**4. TE/TM scheme**

*4.1. Implicit FDTD method based on «transversal electric-transversal magnetic» splitting of the field components in time*

Let us rewrite scheme (5) in an equivalent form

$$\frac{d}{d\tau}\mathbf{u} = \mathbf{T}_0\mathbf{u} + \mathbf{L}\mathbf{v} + \mathbf{j}_u, \quad \frac{d}{d\tau}\mathbf{v} = \mathbf{T}_1\mathbf{v} - \mathbf{L}^T\mathbf{u} + \mathbf{j}_v, \quad (15)$$

where

$$\mathbf{T}_i = \begin{pmatrix} 0 & 0 & -\mathbf{P}_y^i \\ 0 & 0 & \mathbf{P}_x^i \\ \left(\mathbf{P}_y^i\right)^* & -\left(\mathbf{P}_x^i\right)^* & 0 \end{pmatrix}, \quad \mathbf{L} = \begin{pmatrix} 0 & \mathbf{P}_z^0 & 0 \\ -\mathbf{P}_z^1 & 0 & 0 \\ 0 & 0 & 0 \end{pmatrix},$$

$$\mathbf{u} = \begin{pmatrix} \mathbf{h}_x \\ \mathbf{h}_y \\ \mathbf{e}_z \end{pmatrix}, \quad \mathbf{v} = \begin{pmatrix} \mathbf{e}_x \\ \mathbf{e}_y \\ \mathbf{h}_z \end{pmatrix}, \quad \mathbf{j}_u = \begin{pmatrix} 0 \\ 0 \\ -\mathbf{j}_z \end{pmatrix}, \quad \mathbf{j}_v = \begin{pmatrix} -\mathbf{j}_x \\ -\mathbf{j}_y \\ 0 \end{pmatrix}.$$

Applying the TE/TM splitting [1] of the field in time to system (15), the following numerical scheme is obtained

$$\frac{\mathbf{u}^{n+0.5} - \mathbf{u}^{n-0.5}}{\Delta\tau} = \mathbf{T}_0 \frac{\mathbf{u}^{n+0.5} + \mathbf{u}^{n-0.5}}{2} + \mathbf{L}\mathbf{v}^n + \mathbf{j}_u^n, \quad (16)$$

$$\frac{\mathbf{v}^{n+1} - \mathbf{v}^n}{\Delta\tau} = \mathbf{T}_1 \frac{\mathbf{v}^{n+1} + \mathbf{v}^n}{2} - \mathbf{L}^T \mathbf{u}^{n+0.5} + \mathbf{j}_v^{n+0.5}.$$

Scheme (16) is a two-layer scheme

$$\mathbf{B}\frac{\mathbf{y}^{n+1} - \mathbf{y}^n}{\Delta\tau} + \mathbf{A}\mathbf{y}^n = \mathbf{f}^n, \tag{17}$$

where

$$\mathbf{B} = \begin{pmatrix} \mathbf{I} - \alpha\mathbf{T}_0 & \mathbf{0} \\ 2\alpha\mathbf{L}^T & \mathbf{I} - \alpha\mathbf{T}_1 \end{pmatrix}, \qquad \mathbf{A} = \begin{pmatrix} -\mathbf{T}_0 & -\mathbf{L} \\ \mathbf{L}^T & -\mathbf{T}_1 \end{pmatrix},$$

$$\mathbf{y}^n = \begin{pmatrix} \mathbf{u}^{n-0.5} \\ \mathbf{v}^n \end{pmatrix}, \quad \mathbf{f}^n = \begin{pmatrix} \mathbf{j}_u^n \\ \mathbf{j}_v^{n+0.5} \end{pmatrix}, \ \alpha = 0.5\Delta\tau.$$

We study the stability of scheme (17) by the energy inequalities method. Let us take the inner product of both sides in equation (17) with $\mathbf{y}^{n+1} + \mathbf{y}^n$:

$$\langle \mathbf{B}(\mathbf{y}^{n+1} - \mathbf{y}^n), \mathbf{y}^{n+1} + \mathbf{y}^n \rangle + 2\alpha \langle \mathbf{A}\mathbf{y}^n, \mathbf{y}^{n+1} + \mathbf{y}^n \rangle = \langle 2\alpha\mathbf{f}^n, \mathbf{y}^{n+1} + \mathbf{y}^n \rangle.$$

Using the formula

$$\mathbf{y}^n = 0.5\left((\mathbf{y}^{n+1} + \mathbf{y}^n) - (\mathbf{y}^{n+1} - \mathbf{y}^n)\right)$$

we rewrite the last relation in the form

$$\langle [\mathbf{B} - \alpha\mathbf{A}](\mathbf{y}^{n+1} - \mathbf{y}^n), \mathbf{y}^{n+1} + \mathbf{y}^n \rangle + \alpha \langle \mathbf{A}(\mathbf{y}^{n+1} + \mathbf{y}^n), \mathbf{y}^{n+1} + \mathbf{y}^n \rangle = \langle 2\alpha\mathbf{f}^n, \mathbf{y}^{n+1} + \mathbf{y}^n \rangle.$$

The second term on the left hand side is equal to zero since the operator $\mathbf{A}$ is skew-symmetric and, therefore,

$$\langle \mathbf{Q}\mathbf{y}^{n+1}, \mathbf{y}^{n+1} \rangle - \langle \mathbf{Q}\mathbf{y}^n, \mathbf{y}^n \rangle = \langle 2\alpha\mathbf{f}^n, \mathbf{y}^{n+1} + \mathbf{y}^n \rangle$$

where the self-adjointness of the operator $\mathbf{Q} \equiv \mathbf{B} - \alpha\mathbf{A}$ is used.

The last relation allows to prove that the condition

$$\mathbf{Q} \equiv \mathbf{B} - \alpha\mathbf{A} > 0 \ (\mathbf{Q} \text{ is a positively definite matrix})$$

is sufficient for the stability of the scheme.

If the matrix $\mathbf{Q}$ is a positively definite, then we can define a discrete energy as

$$W_{TE/TM}^n = 0.5 \langle \mathbf{Q}\mathbf{y}^n, \mathbf{y}^n \rangle \tag{18}$$

and the discrete *energy conservation law* holds

$$\frac{W_{TE/TM}^{n+1} - W_{TE/TM}^n}{\Delta\tau} = -0.5 \left\langle \left[ \mathbf{e}_x^{n+1} + \mathbf{e}_x^n, \mathbf{e}_y^{n+1} + \mathbf{e}_y^n, \mathbf{e}_z^{n+0.5} + \mathbf{e}_z^{n-0.5} \right], \left[ \mathbf{j}_x^{n+0.5}, \mathbf{j}_y^{n+0.5}, \mathbf{j}_z^n \right] \right\rangle. \tag{19}$$

The stability condition can be rewritten in the form

$$\mathbf{I} + \alpha^2 \mathbf{P}_z^i \left( \mathbf{P}_z^{i*} \right) > 0, \qquad i = 0,1. \tag{21}$$

The last condition resembles the well-known stability condition of the explicit FDTD scheme for one-dimensional problem. The maximal eigenvalue $\lambda_{\max}^i$ of the discrete operator $\mathbf{P}_z^i \left( \mathbf{P}_z^{i*} \right)$ in staircase approximation of the boundary fulfils the relation

$$\lambda_{\max}^i < 4/\Delta z^2, \ i = 0,1,$$

and the stability condition reads

$$\Delta\tau \leq \Delta z. \tag{22}$$

On an equidistant mesh implicit scheme (16) has a second order local approximation error in homogeneous regions, $O(\|\Delta\vec{r}\|^2 + \Delta\tau^2)$, $\Delta\vec{r} = (\Delta x, \Delta y, \Delta z)^T$.

It will be shown in Section 4.5 that, with the time step $\Delta\tau$ equal to the longitudinal mesh step $\Delta z$, scheme (16) does not have dispersion in the longitudinal direction. Relation (22) does not contain information about the transverse mesh. Hence the transverse mesh can be chosen independently from stability considerations.

However, this scheme is implicit and non-economical. The economical scheme modifications based on operator splitting were considered in [1]. In the following we introduce a new explicit scheme with improved dispersion properties in the transverse plane. Due to its explicit character the new scheme is easy parallelizable.

*4.2. Explicit TE/TM scheme in 3D*

Operators $\mathbf{I} - \alpha \mathbf{T}_i$, $i = 0,1$, can be factorized as

$$\mathbf{I} - \alpha \mathbf{T}_i = \mathbf{L}_{\mathbf{T}_i} \mathbf{U}_{\mathbf{T}_i} - \alpha^2 \mathbf{R}_i,$$

where $\mathbf{L}_{\mathbf{T}_i}$ is a low triangular matrix, $\mathbf{U}_{\mathbf{T}_i}$ is an upper triangular one

$$\mathbf{L}_{\mathbf{T}_i} = \begin{pmatrix} \mathbf{I} & 0 & 0 \\ 0 & \mathbf{I} & 0 \\ -\alpha \left(\mathbf{P}_y^i\right)^* & \alpha \left(\mathbf{P}_x^i\right)^* & \mathbf{I} \end{pmatrix}, \qquad \mathbf{U}_{\mathbf{T}_i} = \begin{pmatrix} \mathbf{I} & 0 & \alpha \mathbf{P}_y^i \\ 0 & \mathbf{I} & -\alpha \mathbf{P}_x^i \\ 0 & 0 & \mathbf{I} \end{pmatrix},$$

and the splitting operator is defined as

$$\mathbf{R}_i = \begin{pmatrix} 0 & 0 & 0 \\ 0 & 0 & 0 \\ 0 & 0 & \mathbf{r}_i \end{pmatrix}, \quad \mathbf{r}_i = \left(\mathbf{P}_y^i\right)^* \left(\mathbf{P}_y^i\right) + \left(\mathbf{P}_x^i\right)^* \left(\mathbf{P}_x^i\right), \; i = 0,1.$$

Hence, operator $\mathbf{B}$ can be written as

$$\mathbf{B} = \mathbf{L}_{\mathbf{L}} + \mathbf{L}_{\mathbf{T}} \mathbf{U}_{\mathbf{T}} - \alpha^2 \mathbf{R},$$

where

$$\mathbf{L}_{\mathbf{L}} = \begin{pmatrix} 0 & 0 \\ 2\alpha \mathbf{L}^* & 0 \end{pmatrix}, \qquad \mathbf{L}_{\mathbf{T}} = \begin{pmatrix} \mathbf{L}_{\mathbf{T}_0} & 0 \\ 0 & \mathbf{L}_{\mathbf{T}_1} \end{pmatrix}, \qquad \mathbf{U}_{\mathbf{T}} = \begin{pmatrix} \mathbf{U}_{\mathbf{T}_0} & 0 \\ 0 & \mathbf{U}_{\mathbf{T}_1} \end{pmatrix}, \qquad \mathbf{R} = \begin{pmatrix} \mathbf{R}_0 & 0 \\ 0 & \mathbf{R}_1 \end{pmatrix}.$$

Neglecting the term $\alpha^2 \mathbf{R}$ of the order $O(\Delta \tau^2)$ the implicit scheme (16) can be reduced to the explicit one

$$\mathbf{B}^e \frac{\mathbf{y}^{n+1} - \mathbf{y}^n}{\Delta \tau} + \mathbf{A} \mathbf{y}^n = \mathbf{f}^n, \qquad \mathbf{B}^e = \mathbf{L}_{\mathbf{L}} + \mathbf{L}_{\mathbf{T}} \mathbf{U}_{\mathbf{T}}. \tag{23}$$

The following relations hold

$$\mathbf{A} = -\mathbf{A}^*, \qquad \mathbf{Q}^e = \mathbf{Q}^{e*}, \qquad \mathbf{Q}^e = \mathbf{B}^e - \alpha \mathbf{A}.$$

On equidistant mesh *explicit* scheme (23) has a second order local approximation error in homogeneous regions, $O(\|\Delta \vec{r}\|^2 + \Delta \tau^2)$, $\Delta \vec{r} = (\Delta x, \Delta y, \Delta z)^T$.

The explicit scheme can be rewritten as

$$\mathbf{L}_{\mathbf{T}_0} \mathbf{U}_{\mathbf{T}_0} \frac{\mathbf{u}^{n+0.5} - \mathbf{u}^{n-0.5}}{\Delta \tau} = \mathbf{T}_0 \mathbf{u}^{n-0.5} + \mathbf{L} \mathbf{v}^n + \mathbf{j}_u^n,$$

$$\mathbf{L}_{\mathbf{T}_1} \mathbf{U}_{\mathbf{T}_1} \frac{\mathbf{v}^{n+1} - \mathbf{v}^n}{\Delta \tau} = \mathbf{T}_1 \mathbf{v}^n - \mathbf{L}^* \mathbf{u}^{n+0.5} + \mathbf{j}_v^{n+0.5},$$

$$\mathbf{u}^{n+0.5} = \begin{pmatrix} h_x^{n+0.5} \\ h_y^{n+0.5} \\ e_z^{n+0.5} \end{pmatrix}, \qquad \mathbf{v}^n = \begin{pmatrix} e_x^n \\ e_y^n \\ h_z^n \end{pmatrix}.$$

In the original variables and in detailed notation it reads

$$\widehat{\mathbf{h}}_x^n = \widehat{\mathbf{h}}_x^{n-0.5} + \alpha \mathbf{M}_{\mu_x^{-1}} \left[ \mathbf{P}_z \widehat{\mathbf{e}}_y^n - \mathbf{P}_y \widehat{\mathbf{e}}_z^{n-0.5} \right],$$

$$\widehat{\mathbf{h}}_y^n = \widehat{\mathbf{h}}_y^{n-0.5} + \alpha \mathbf{M}_{\mu_y^{-1}} \left[ -\mathbf{P}_z \widehat{\mathbf{e}}_x^n + \mathbf{P}_x \widehat{\mathbf{e}}_z^{n-0.5} \right],$$

$$\widehat{\mathbf{e}}_z^{n+0.5} = \widehat{\mathbf{e}}_z^{n-0.5} + 2\alpha \mathbf{M}_{\varepsilon_z^{-1}} \left[ \mathbf{P}_y^* \widehat{\mathbf{h}}_x^n - \mathbf{P}_x^* \widehat{\mathbf{h}}_y^n + \widehat{\overline{\mathbf{j}}}_z^n \right],$$

$$\widehat{\mathbf{h}}_x^{n+0.5} = \widehat{\mathbf{h}}_x^n + \alpha \mathbf{M}_{\mu_x^{-1}} \left[ \mathbf{P}_z \widehat{\mathbf{e}}_y^n - \mathbf{P}_y \widehat{\mathbf{e}}_z^{n+0.5} \right],$$

$$\widehat{\mathbf{h}}_y^{n+0.5} = \widehat{\mathbf{h}}_y^n + \alpha \mathbf{M}_{\mu_y^{-1}} \left[ -\mathbf{P}_z \widehat{\mathbf{e}}_x^n + \mathbf{P}_x \widehat{\mathbf{e}}_z^{n+0.5} \right] \quad ;$$

$$\widehat{\mathbf{e}}_x^{n+0.5} = \widehat{\mathbf{e}}_x^n + \alpha \mathbf{M}_{\varepsilon_x^{-1}} \left[ \mathbf{P}_z^* \widehat{\mathbf{h}}_y^{n+0.5} - \mathbf{P}_y^* \widehat{\mathbf{h}}_z^n + \widehat{\overline{\mathbf{j}}}_x^{n+0.5} \right],$$

$$\widehat{\mathbf{e}}_y^{n+0.5} = \widehat{\mathbf{e}}_y^n + \alpha \mathbf{M}_{\varepsilon_y^{-1}} \left[ -\mathbf{P}_z^* \widehat{\mathbf{h}}_x^{n+0.5} + \mathbf{P}_x^* \widehat{\mathbf{h}}_z^n + \widehat{\overline{\mathbf{j}}}_y^{n+0.5} \right],$$

$$\mathbf{h}_z^{n+1} = \mathbf{h}_z^n + 2\alpha \mathbf{M}_{\mu_z^{-1}} \left[ \mathbf{P}_y \widehat{\mathbf{e}}_x^{n+0.5} - \mathbf{P}_x \widehat{\mathbf{e}}_y^{n+0.5} \right],$$

$$\widehat{\mathbf{e}}_x^{n+1} = \widehat{\mathbf{e}}_x^{n+0.5} + \alpha \mathbf{M}_{\varepsilon_x^{-1}} \left[ \mathbf{P}_z^* \widehat{\mathbf{h}}_y^{n+0.5} - \mathbf{P}_y^* \widehat{\mathbf{h}}_z^{n+1} + \widehat{\overline{\mathbf{j}}}_x^{n+0.5} \right],$$

$$\widehat{\mathbf{e}}_y^{n+1} = \widehat{\mathbf{e}}_y^{n+0.5} + \alpha \mathbf{M}_{\varepsilon_y^{-1}} \left[ -\mathbf{P}_z^* \widehat{\mathbf{h}}_x^{n+0.5} + \mathbf{P}_x^* \widehat{\mathbf{h}}_z^{n+1} + \widehat{\overline{\mathbf{j}}}_y^{n+0.5} \right],$$

where we have introduced the auxiliary variables $\widehat{\mathbf{h}}_x^n, \widehat{\mathbf{h}}_y^n, \widehat{\mathbf{e}}_x^{n+0.5}, \widehat{\mathbf{e}}_y^{n+0.5}$. Hence we have 10 update equations at each time step of the same form as in Yee's scheme and the numerical effort at each time step is scaled as 5/3 compared to Yee's algorithm. But it follows from Eq.(25) that the explicit TE/TM scheme allows a larger time step and for the mesh $\Delta x = \Delta y = \sqrt{2} \Delta z$ the new scheme requires only ~18% more computational time than Yee's scheme. The memory requirements of the new scheme are the same as for Yee's scheme: only one vector for each field component is required.

*4.3. The explicit TE/TM for rotationally symmetric geometries*

For rotationally symmetric cases the Maxwell equations are reduced to equations for azimuthal harmonics. The azimuthal derivative is taken explicitly and we can use another modified operator

$$\mathbf{B}^{e,r} = \mathbf{B} + \alpha^2 \mathbf{R}^r \quad ,$$

$$\mathbf{R}^r = \begin{pmatrix} \mathbf{R}_0^r & 0 \\ 0 & \mathbf{R}_1^r \end{pmatrix}, \quad \mathbf{R}_i^r = \begin{pmatrix} 0 & 0 & 0 \\ 0 & 0 & 0 \\ 0 & 0 & \mathbf{r}_i^r \end{pmatrix}, \quad \mathbf{r}_i^r = \left( \mathbf{P}_r^i \right)^* \left( \mathbf{P}_r^i \right).$$

The explicit scheme

$$\mathbf{B}^{er} \frac{\mathbf{y}^{n+1} - \mathbf{y}^n}{\Delta \tau} + \mathbf{A} \mathbf{y}^n = \mathbf{f}^n \tag{24}$$

at mode $m$ reads ($\mathbf{j}_v \equiv 0$)

$$\widehat{\mathbf{h}}_\varphi^n = \widehat{\mathbf{h}}_\varphi^{n-0.5} + \alpha \mathbf{M}_{\mu_\varphi^{-1}} \left[ \mathbf{P}_z \widehat{\mathbf{e}}_r^n - \mathbf{P}_r \widehat{\mathbf{e}}_z^{n-0.5} \right],$$

$$\widehat{\mathbf{h}}_r^n = \widehat{\mathbf{h}}_r^{n-0.5} + \alpha \mathbf{M}_{\mu_r^{-1}} \left[ -\mathbf{P}_z \widehat{\mathbf{e}}_\varphi^n + m \widehat{\mathbf{e}}_z^{n-0.5} \right],$$

$$\widehat{\mathbf{e}}_z^{n+0.5} = \widehat{\mathbf{e}}_z^{n-0.5} + \left( \mathbf{I} + \alpha^2 m^2 \mathbf{M}_{\mu_z^{-1}} \mathbf{M}_{\varepsilon_r^{-1}} \right)^{-1} 2\alpha \mathbf{M}_{\varepsilon_z^{-1}} \left[ \mathbf{P}_r^* \widehat{\mathbf{h}}_\varphi^n - m \widehat{\mathbf{h}}_r^n + \widehat{\overline{\mathbf{j}}}_z^n \right],$$

$$\widehat{\mathbf{h}}_\varphi^{n+0.5} = \widehat{\mathbf{h}}_\varphi^n + \alpha \mathbf{M}_{\mu_\varphi^{-1}} \left[ \mathbf{P}_z \widehat{\mathbf{e}}_r^n - \mathbf{P}_r \widehat{\mathbf{e}}_z^{n+0.5} \right],$$

$$\widehat{\mathbf{h}}_r^{n+0.5} = \widehat{\mathbf{h}}_r^n + \alpha \mathbf{M}_{\mu_r^{-1}} \left[ -\mathbf{P}_z \widehat{\mathbf{e}}_\varphi^n + m \widehat{\mathbf{e}}_z^{n+0.5} \right];$$

$$\widehat{\mathbf{e}}_\varphi^{n+0.5} = \widehat{\mathbf{e}}_\varphi^n + \alpha \mathbf{M}_{\varepsilon_\varphi^{-1}} \left[ \mathbf{P}_z^* \widehat{\mathbf{h}}_r^{n+0.5} - \mathbf{P}_r^* \widehat{\mathbf{h}}_z^n + \widehat{\overline{\mathbf{j}}}_\varphi^{n+0.5} \right],$$

$$\widehat{\mathbf{e}}_r^{n+0.5} = \widehat{\mathbf{e}}_r^n + \alpha \mathbf{M}_{\varepsilon_r^{-1}} \left[ -\mathbf{P}_z^* \widehat{\mathbf{h}}_\varphi^{n+0.5} + m \widehat{\mathbf{h}}_z^n + \widehat{\overline{\mathbf{j}}}_r^{n+0.5} \right],$$

$$\mathbf{h}_z^{n+1} = \mathbf{h}_z^n + \left( \mathbf{I} + \alpha^2 m^2 \mathbf{M}_{\varepsilon_z^{-1}} \mathbf{M}_{\mu_r^{-1}} \right)^{-1} 2\alpha \mathbf{M}_{\mu_z^{-1}} \left[ \mathbf{P}_r \widehat{\mathbf{e}}_\varphi^{n+0.5} - m \widehat{\mathbf{e}}_r^{n+0.5} \right],$$

$$\widehat{\mathbf{e}}_\varphi^{n+1} = \widehat{\mathbf{e}}_\varphi^{n+0.5} + \alpha \mathbf{M}_{\varepsilon_\varphi^{-1}} \left[ \mathbf{P}_z^* \widehat{\mathbf{h}}_r^{n+0.5} - \mathbf{P}_r^* \widehat{\mathbf{h}}_z^{n+1} + \widehat{\overline{\mathbf{j}}}_\varphi^{n+0.5} \right],$$

$$\widehat{\mathbf{e}}_r^{n+1} = \widehat{\mathbf{e}}_r^{n+0.5} + \alpha \mathbf{M}_{\varepsilon_r^{-1}} \left[ -\mathbf{P}_z^* \widehat{\mathbf{h}}_\varphi^{n+0.5} + m \widehat{\mathbf{h}}_z^{n+1} + \widehat{\overline{\mathbf{j}}}_r^{n+0.5} \right].$$

If the material matrices $\mathbf{M}_{\mu^{-1}}$, $\mathbf{M}_{\varepsilon^{-1}}$ are diagonal the new scheme is explicit. On an equidistant mesh *explicit* scheme (24) has a second order local approximation error in homogeneous regions, $O(\|\Delta \vec{r}\|^2 + \Delta \tau^2)$, $\Delta \vec{r} = (\Delta r, \Delta z)^T$.

In order to reach the maximal time step and to avoid non-diagonal material matrices we use the Simplified Conformal method introduced in [14].

*4.4 Energy conservation and stability*

Energy conservation and stability condition of the implicit TE/TM scheme (8) was considered above (see [1]). Following the same way the discrete energy for the explixit TE/TM scheme can be defined by the relation

$$W_{TE/TM}^{n,e} = 0.5 \langle \mathbf{Q}^e \mathbf{y}^n, \mathbf{y}^n \rangle, \qquad \mathbf{Q}^e = \mathbf{B}^e - \alpha \mathbf{A}.$$

and a discrete *energy conservation law* in form (19) holds.

The sufficient stability condition of positive definiteness of the matrix $\mathbf{Q}^e$,

$$\mathbf{Q}^e > 0,$$

can be rewritten in the form

$$\begin{cases} \mathbf{I} + \alpha^2 \mathbf{P}_z^i \left( \mathbf{P}_z^{i*} \right) > 0, \\ \mathbf{I} + \alpha^2 \left[ \mathbf{P}_x^i \left( \mathbf{P}_x^{i*} \right) + \mathbf{P}_y^i \left( \mathbf{P}_y^{i*} \right) \right] > 0, \end{cases} \quad i = 0,1.$$

From consideration of the maximal eigenvalue of the discrete Laplace operator $\mathbf{P}_x^i \left( \mathbf{P}_x^{i*} \right) + \mathbf{P}_y^i \left( \mathbf{P}_y^{i*} \right)$ it follows that for a vacuum domain with a *staircase* approximation of the boundary the stability condition reads

$$\begin{cases} \Delta \tau \leq \Delta z, \\ \Delta \tau \leq \left( \Delta x^{-2} + \Delta y^{-2} \right)^{-0.5}. \end{cases} \tag{25}$$

For a rotationally symmetric case it reduces to the form

$$\Delta \tau \leq \min(\Delta z, \Delta r).$$

Note that the last stability condition does not include the azimuthal mode number $m$. For comparison, Yee's scheme requires to reduce the stable time step [10]

$$\Delta \tau \leq \frac{1}{\sqrt{(1.2 + 0.4 m^2) \Delta r^{-2} + 1.2 \Delta z^{-2}}}$$

for higher azimuthal modes, that increases the computational effort considerably.

## 4.5 Dispersion relation of the TE/TM scheme

Following the conventional procedure [12] the dispersion relations can be obtained in the form

$$\frac{\sin^2 \Omega}{\Delta \tau^2} = \frac{\sin^2 K_z}{\Delta z^2} + \left( \frac{\sin^2 K_x}{\Delta x^2} + \frac{\sin^2 K_y}{\Delta y^2} \right) \cos^2 \Omega,$$

for the implicit TE/TM scheme and in the form

$$\frac{\sin^2 \Omega}{\Delta \tau^2} = \frac{\sin^2 K_z}{\Delta z^2} + \left( \frac{\sin^2 K_x}{\Delta x^2} + \frac{\sin^2 K_y}{\Delta y^2} \right)\left(1 - \frac{\Delta \tau^2}{\Delta z^2} \sin^2 K_z \right), \quad (27)$$

for the explicit one. Here $\Omega = 0.5\omega \Delta \tau / c$, $K_x = 0.5 k_x \Delta x$, $K_y = 0.5 k_y \Delta y$, $K_z = 0.5 k_z \Delta z$.

With the "magic" time step $\Delta \tau = \Delta z$ both schemes do not have dispersion in the longitudinal direction.

The explicit scheme can have yet two directions in the transverse $XY$ plane with zero dispersion. Indeed for equal transverse mesh steps

$$\Delta x = \Delta y = h,$$

the dispersion relation for transverse plane waves ($K_z = 0$) reads

$$\sin^2 \Omega = \frac{\Delta \tau^2}{h^2} \left( \sin^2 K_x + \sin^2 K_y \right)$$

and has the same form as the dispersion relation for 2D Yee's scheme.

For the implicit scheme the dispersion relation for transverse waves reads

$$\tan^2 \Omega = \frac{\Delta \tau^2}{h^2} \left( \sin^2 K_x + \sin^2 K_y \right).$$

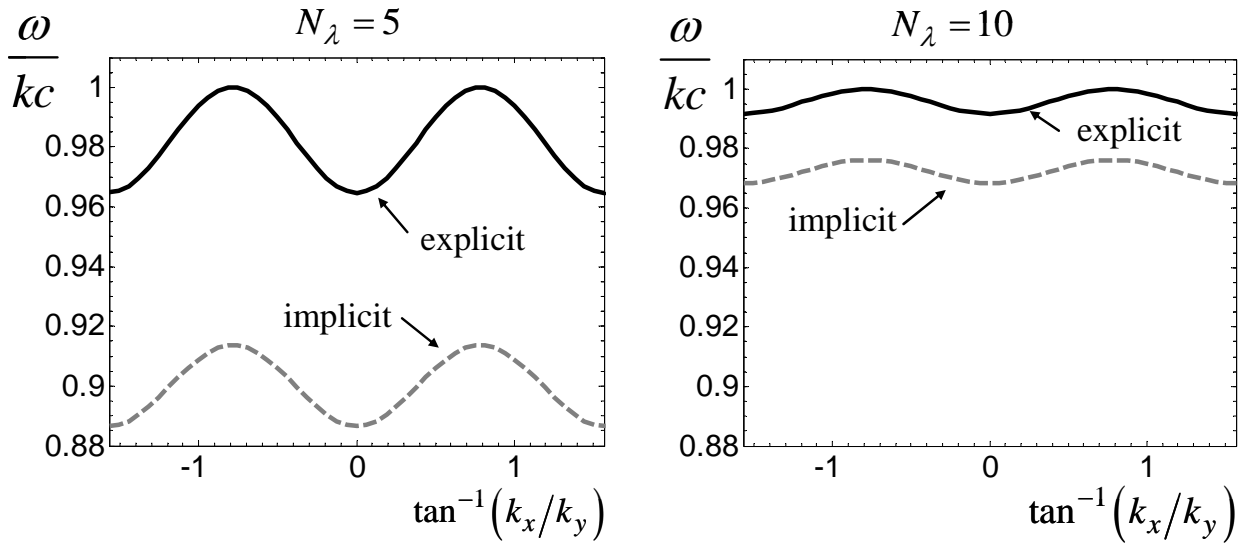

Fig. 2. Comparison of the dispersion curves in the transverse plane.

Fig. 2 shows the dispersion curves in the transverse plane for the time step $\Delta \tau = h/\sqrt{2}$ and for different mesh resolutions $N_\lambda = \lambda / h$, where $\lambda$ is the wave length.

The dispersion relation (27) means that Eq.(25) is a necessary stability condition for the explicit TE/TM scheme (23).

*4.5 Charge conservation*

In this section we consider a discrete analogue of the charge conservation law (2). Let us introduce a discrete divergence matrix

$$\mathbf{S} = \begin{pmatrix} \mathbf{S}_{11} & \mathbf{S}_{12} \\ \mathbf{S}_{21} & \mathbf{S}_{22} \end{pmatrix} \equiv \begin{pmatrix} \mathbf{P}_x^h & \mathbf{P}_y^h & 0 & 0 & 0 & \mathbf{P}_z^h \\ 0 & 0 & \mathbf{P}_z^e & \mathbf{P}_x^e & \mathbf{P}_y^e & 0 \end{pmatrix}$$

and multiply Eq. (23) with it. We obtain

$$\mathbf{SB}\frac{\mathbf{y}^{n+1} - \mathbf{y}^n}{\Delta \tau} = \mathbf{Sf}^n, \tag{28}$$

where the equality $\mathbf{SA} = \mathbf{0}$ was used. The last equation can be considered as a charge conservation law for the discrete charge $\mathbf{q}^n = \mathbf{SBy}^n$. However, the matrix product

$$\mathbf{SB} = \begin{pmatrix} \mathbf{S}_{11} & \mathbf{S}_{12}(\mathbf{I} - \alpha \mathbf{T}_0) \\ \mathbf{S}_{21}(\mathbf{I} + \alpha \mathbf{T}_1) & \mathbf{S}_{22} \end{pmatrix}$$

differs from matrix $\mathbf{S}$ by terms of the order $O(\Delta \tau)$. Hence relation (28) is only a first order approximation of the continuity equation (2). In the following we will show that a second order approximation of the continuity equation (2) holds.

Let us introduce new field and current vectors

$$\mathbf{z}^n = \begin{pmatrix} \mathbf{u}^{n+0.5} \\ \mathbf{v}^n \end{pmatrix}, \quad \overline{\mathbf{f}}^n = \begin{pmatrix} \mathbf{j}_u^{n+1} \\ \mathbf{j}_v^{n+0.5} \end{pmatrix},$$

and rewrite the scheme (23) in the form

$$\left(-\mathbf{B}^e\right)^* \frac{\mathbf{z}^{n+1} - \mathbf{z}^n}{\Delta \tau} + \mathbf{A}\mathbf{z}^n = \overline{\mathbf{f}}^n, \tag{29}$$

Let us build two sums of equations (23) and (29)

$$\mathbf{B}^e \frac{\mathbf{y}^{n+1} - \mathbf{y}^n}{\Delta \tau} - \left(\mathbf{B}^e\right)^* \frac{\mathbf{z}^{n+1} - \mathbf{z}^n}{\Delta \tau} + \mathbf{A}(\mathbf{y}^n + \mathbf{z}^n) = \mathbf{f}^n + \overline{\mathbf{f}}^n, \tag{30}$$

$$\mathbf{B}^e \frac{\mathbf{y}^{n+1} - \mathbf{y}^n}{\Delta \tau} - \left(\mathbf{B}^e\right)^* \frac{\mathbf{z}^n - \mathbf{z}^{n-1}}{\Delta \tau} + \mathbf{A}(\mathbf{y}^n + \mathbf{z}^{n-1}) = \mathbf{f}^n + \overline{\mathbf{f}}^{n-1} \tag{31}.$$

If we apply the operator $\mathbf{S}$ to these equations and use the equality

$$\mathbf{B}^e - \left(\mathbf{B}^e\right) = 2\mathbf{I} + 2\alpha \mathbf{A} + \alpha^2 \mathbf{R},$$

then we obtain

$$\frac{\mathbf{S}(\mathbf{I} + \alpha^2 \mathbf{R})}{\Delta \tau}\left[\begin{pmatrix} \mathbf{u}^{n+1.5} + \mathbf{u}^{n+0.5} \\ 2\mathbf{v}^{n+1} \end{pmatrix} - \begin{pmatrix} \mathbf{u}^{n+0.5} + \mathbf{u}^{n-0.5} \\ 2\mathbf{v}^n \end{pmatrix}\right] - \frac{\alpha}{\Delta \tau}\begin{pmatrix} \mathbf{0} \\ \mathbf{S}_{21}\mathbf{T}_0(\mathbf{u}^{n+1.5} - 2\mathbf{u}^{n+0.5} + \mathbf{u}^{n-0.5}) \end{pmatrix} = \mathbf{S}(\mathbf{f}^n + \overline{\mathbf{f}}^n), \tag{32}$$

$$\frac{\mathbf{S}(\mathbf{I} + \alpha^2 \mathbf{R})}{\Delta \tau}\left[\begin{pmatrix} 2\mathbf{u}^{n+0.5} \\ \mathbf{v}^{n+1} + \mathbf{v}^n \end{pmatrix} - \begin{pmatrix} 2\mathbf{u}^{n-0.5} \\ \mathbf{v}^n + \mathbf{v}^{n-1} \end{pmatrix}\right] - \frac{\alpha}{\Delta \tau}\begin{pmatrix} \mathbf{S}_{12}\mathbf{T}_1(\mathbf{v}^{n+1} - 2\mathbf{v}^n + \mathbf{v}^{n-1}) \\ \mathbf{0} \end{pmatrix} = \mathbf{S}(\mathbf{f}^n + \overline{\mathbf{f}}^{n-1}). \tag{33}.$$

The first row of the matrix in Eq. (32) gives

$$\frac{\overline{\mathbf{q}}_h^{n+1} - \overline{\mathbf{q}}_h^n}{\Delta \tau} = \mathbf{0}, \tag{34}$$

$$\overline{\mathbf{q}}_h^n = \mathbf{P}_x^h \frac{\mathbf{h}_x^{n+0.5} + \mathbf{h}_x^{n-0.5}}{2} + \mathbf{P}_y^h \frac{\mathbf{h}_y^{n+0.5} + \mathbf{h}_y^{n-0.5}}{2} + \mathbf{P}_z^h (\mathbf{I} + \alpha^2 \mathbf{r}_1)\mathbf{h}_z^n.$$

The second row of the matrix in Eq. (33) gives

$$\frac{\overline{\mathbf{q}}_e^{n+0.5} - \overline{\mathbf{q}}_e^{n-0.5}}{\Delta \tau} + \mathbf{S}_e \overline{\mathbf{j}}^n = \mathbf{0}, \tag{35}$$

$$\overline{\mathbf{q}}_e^{n+0.5} = \mathbf{P}_x^e \frac{\mathbf{e}_x^{n+1} + \mathbf{e}_x^n}{2} + \mathbf{P}_y^e \frac{\mathbf{e}_y^{n+1} + \mathbf{e}_y^n}{2} + (\mathbf{I} + \alpha^2 \mathbf{r}_0)\mathbf{P}_z^e \mathbf{e}_z^{n+0.5},$$

$$\overline{\mathbf{j}}^n = \left[ 0.5\left(\mathbf{j}_x^{n+0.5} + \mathbf{j}_x^{n-0.5}\right), 0.5\left(\mathbf{j}_y^{n+0.5} + \mathbf{j}_y^{n-0.5}\right), \mathbf{j}_z^n \right]^T.$$

Eq. (35) is a second order approximation of the continuity equation (2).

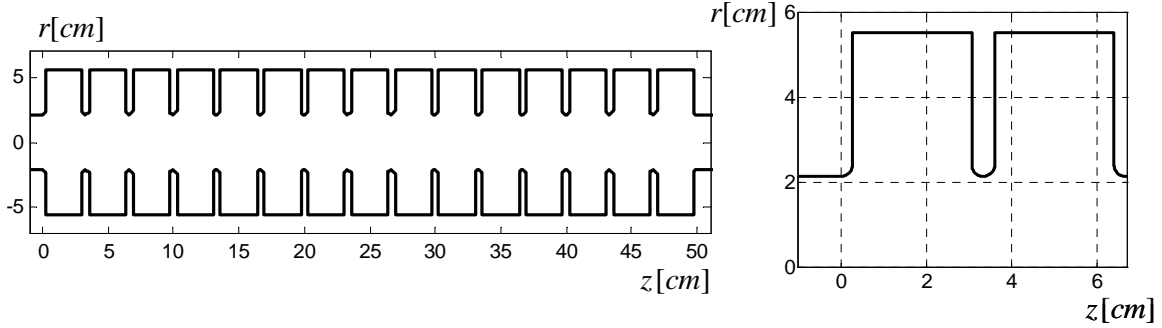

Fig. 3. The geometry of a transverse deflecting cavity. The left figure shows the whole structure of about 50 cm length. The right figure describes a geometry of the first 2 cells.

## 5. Numerical tests

The advantage of the implicit TE/TM compared to Yee's scheme was shown in [1]. Hence here we only show that the explicit scheme gives equally accurate results.

Our first example is a transverse deflecting structure (TDS) to be used in a new Free Electron Laser project at Deutsches Elektronen Synchrotron [15]. The geometry dimensions of TDS in rotationally symmetric approximation are given in Fig. 3. For the Gaussian bunch with RMS width $\sigma = 300 \ \mu m$ we have calculated a dipole transverse wake potential [13]

$$W_\perp^1(s) = \left|\mathbf{W}_\perp^1(s, r, \theta = 0)\right| r^{-1}, \quad \mathbf{W}_\perp^1(s, r, \theta) = \frac{1}{Q} \int_{-\infty}^{\infty} \left[\mathbf{E}_\perp^1 + (\vec{c} \times \mathbf{B}^1)_\perp\right]_{t=(z+s)/v} dz$$

with the implicit (16) and the explicit (24) schemes for rotationally symmetric structures.

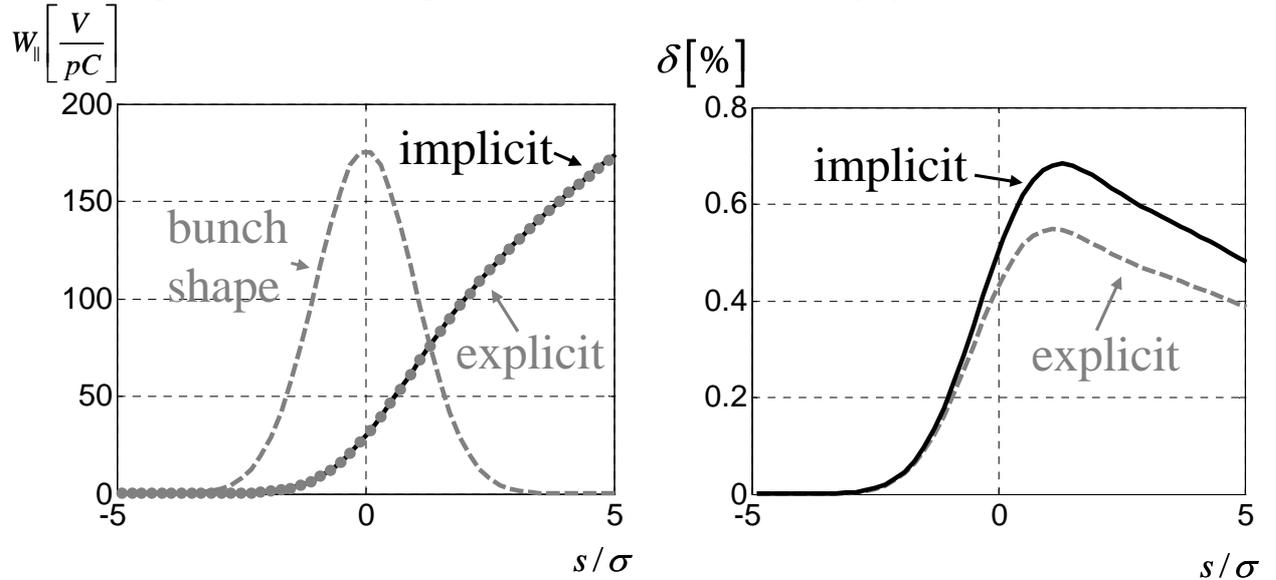

Fig. 4. Comparison of the transverse wake potentials of a TDS structure obtained with the explicit and the implicit schemes for the Gaussian bunch with RMS width of 0.3 mm.

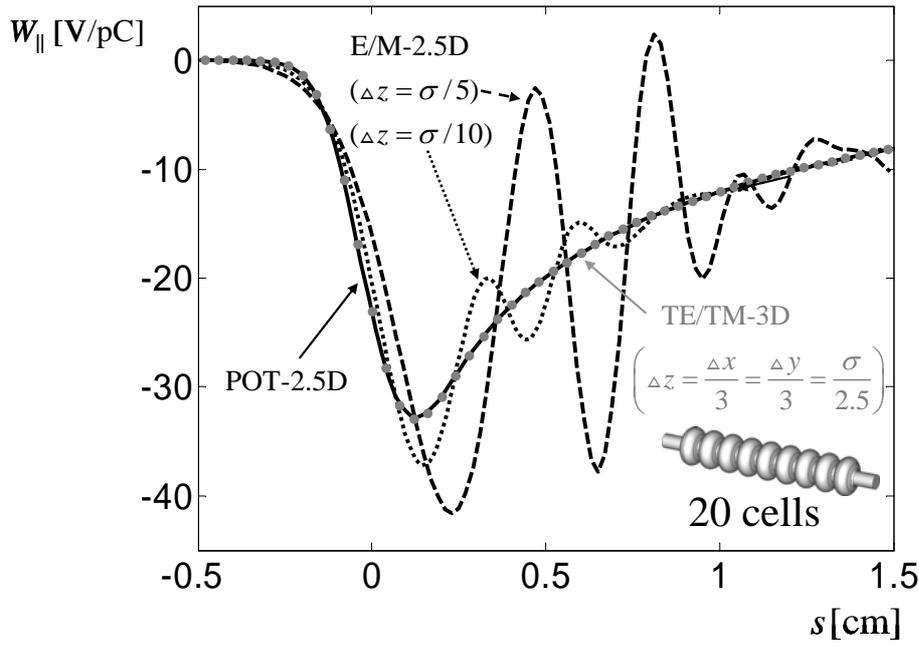

Fig. 5. Comparison of the wake potentials obtained by different methods for the structure consisting of 20 TESLA cells excited by a Gaussian bunch with $\sigma = 1$ mm. The solid line shows the reference solution obtained with the help of the scheme described in [5]. The dashed line describes the solution obtained by classical Yee's scheme with a mesh resolution of 5 mesh steps per $\sigma$. The dotted line describes the solution obtained by Yee's scheme with two times denser resolution in the longitudinal direction. The picture shows a coincidence of the reference result (solid line) with the results on the coarse mesh obtained from the implicit 3D TE/TM code (gray points).

Fig. 4 shows a comparison of the results. We have repeated the calculation with the implicit method on a fine mesh ($\sigma/\Delta z = \sigma/\Delta r = 10$) and compared the result with the results obtained by the explicit and the implicit schemes on the coarse mesh ($\sigma/\Delta z = \sigma/\Delta r = 5$). Fig. 4 on the right shows the relative error

$$\delta = \left|W_\|^{coarse} - W_\|^{fine}\right| / \left(\max W_\|^{fine} - \min W_\|^{fine}\right) \cdot 100\%$$

between the wakes calculated with the implicit TE/TM scheme (16) on the fine mesh and the wakes calculated with the explicit (24) and implicit (16) TE/TM schemes on the coarse mesh.

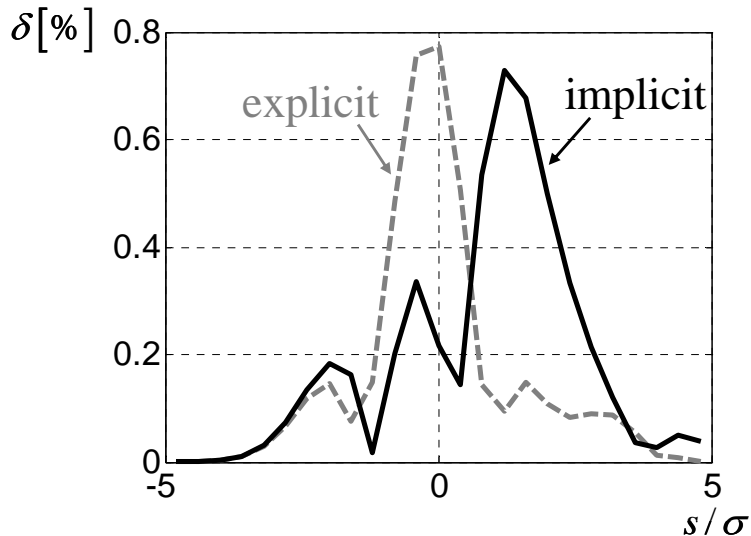

Fig. 6. The error in the wakepotential for the explicit and the implicit TE/TM schemes.

As a next example we consider a structure consisting of 20 TESLA cells (see [15] for the exact geometry of the cavity cell) bounded by infinite ingoing and outgoing pipes with 35 mm diameter. We use the explicit three dimensional scheme (23) for this example.

Fig. 5 shows the longitudinal wake potential [13]

$$W_{\parallel}(s,x,y) = -\frac{1}{Q}\int_{-\infty}^{\infty}\left[E_z(x,y,z,t)\right]_{t=(z+s)/v}dz$$

for a Gaussian bunch with a RMS length of $\sigma = 1$ mm moving on the axis. The solid line (POT-2.5D) corresponds to the accurate reference solution obtained with the vector potential method [4]. The two other lines show results obtained with different mesh resolutions from the TBCI code [16], based on the classical Yee's scheme (E/M-2.5D). The oscillations, that appear, are due to the dispersion error of the Yee's scheme. The gray points represent the result obtained by the implicit three dimensional scheme (16) (marked as TE/TM-3D).

We have repeated the calculation with the explicit method (23) and compared the result with the accurate reference 2D simulation result (POT-2.5D). Fig. 6 shows the relative error

$$\delta = \left|W_{\parallel}^{2D} - W_{\parallel}^{3D}\right|/\left(\max W_{\parallel}^{2D} - \min W_{\parallel}^{2D}\right)\cdot 100\%$$

between the wakes calculated with 2D scheme (POT-2.5D) on a very fine mesh and the wakes calculated with 3D TE/TM schemes (16), (23) on the coarse mesh.

It can be seen that the explicit three dimensional TE/TM scheme produces equally accurate results as the earlier introduced implicit TE/TM scheme. The explicit TE/TM scheme does not suffer from the numerical dispersion and the mesh can be chosen quite coarse. Indeed, the three dimensional code uses only 2.5 mesh points per $\sigma$ in the longitudinal direction. In the transverse direction the mesh steps are even three times larger.

## 6. Conclusion

An explicit scheme for the calculation of electromagnetic fields in the vicinity of relativistic charged bunches was introduced. As shown by several numerical examples the scheme is able to model curved boundaries with a high accuracy and allows for a non-deteriorating calculation of the field solution for very long simulation times. Due to its explicit character the new scheme is faster than the earlier introduced implicit one and it can be easily parallelized.

## References

[1] I. Zagorodnov, T. Weiland, TE/TM scheme for computation of electromagnetic fields in accelerators, J. Comput. Phys. 207 (2005) 69.
[2] A. N. Novokhatski, The computer code NOVO for the calculation of wake potentials of very short ultra-relativistic bunches, SLAC–PUB–11556, 2005.
[3] A. Pukhov, 3D Electromagnetic Relativistic Particle-In-Cell Code VLPL: Virtual Laser Plasma Lab, J. Plasma Physics, 61 (1999) 425.
[4] R. Hampel, I. Zagorodnov, T.Weiland, New discretization scheme for wake field computation in cylindrically symmetric structures, in: Proc. of the EPAC 2004, Lucerne, p. 2556.
[5] I. Zagorodnov, R. Schuhmann, T. Weiland, Long-time computation of electromagnetic fields in the vicinity of a relativistic source, J. Comput. Phys. 191 (2003) 525.
[6] T. Lau, E. Gjonaj, T. Weiland, Time Integration Methods for Particle Beam Simulations with the Finite Integration Theory, Frequenze, 59 (2005) 210.
[7] M. Kärkkäinen et al, Low-Dispersion Wake Field Calculation Tools, in: Proc. of the

ICAP 2006, Chamonix Mont-Blanc , p. 35.
[8]  K.S. Yee, Numerical solution of initial boundary value problems involving Maxwell's equations in isotropic media, IEEE Trans. Antennas and Propagation 14 (1966) 302.
[9]  I. Zagorodnov, R. Schuhmann, T. Weiland, A uniformly stable conformal FDTD-method on Cartesian grids, Int. J. Numer. Model. 16 (2003) 127.
[10] Y.H. Chin, User's guide for ABCI Version 8.7, CERN-SL-94-02-AP, CERN, 1994.
[11] T. Weiland, Time domain electromagnetic field computation with finite difference methods, Int. J. Numer. Model. 9 (1996) 295.
[12] A. Taflove, S.C.Hagness, Ed., Computational Electrodynamics: The Finite-Difference Time-Domain Method, Artech House, London, 2000.
[13] A.C. Chao, Physics of Collective Beam Instabilities in High Energy Accelerators, John Wiley& Sons, New York, 1993.
[14] Zagorodnov I., Schuhmann R., Weiland T., Conformal FDTD-methods to avoid Time Step Reduction with and without Cell Enlargement, J. Comput. Phys. 225 (2007) 1493 .
[15] TESLA Technical Design Report, DESY 2001-011, Hamburg, Germany, 2001, Part II.
[16] T. Weiland, TBCI and URMEL - New computer codes for wake field and cavity mode calculations, IEEE Trans. Nuclear Science, 30 (1983) 2489.